# From Fabrication Disorder to Quantum-Protocol Yield in Inverse-Designed Silicon Photonic Beam Splitters

Naïl Letty[1], Tahmeed Ul Islam[1], Wayesh Qarony[*,1,2,3]

[1]*Department of Electrical and Computer Engineering, University of Central Florida, Orlando, FL 32816, USA*

[2]*Department of Physics, University of Central Florida, Orlando, FL 32816, USA*

[3]*CREOL, The College of Optics and Photonics, University of Central Florida, Orlando, FL 32816, USA*

* *Corresponding author:* wayesh@ucf.edu

**ABSTRACT.** Telecom-band silicon spin–photon qubits offer a promising platform for scalable quantum networks and distributed information processing, but their fixed emission wavelengths require compact, low-loss beam splitters with precise splitting and fabrication robustness. Conventional designs can face trade-offs between footprint, optical performance, and wavelength-specific operation. Here, we employ adjoint-based inverse design to optimize photonic geometries beyond conventional layouts, developing 1×2 Y-splitters and 2×2 directional couplers for the silicon T (1326 nm) and AI1 (1482.44 nm) spin–photon qubits. The devices achieve footprints of 2.9 and 26 μm², respectively, with excess losses of 0.14–0.44 dB and back-reflection below $5 \times 10^{-4}$. We assess fabrication tolerance using 100-realization Monte Carlo simulations incorporating etch bias, line-edge roughness, thickness, and refractive-index variations. Mirror symmetry reduces the splitting-ratio spread of the Y-splitter by a factor of seven to nine relative to the directional coupler. We translate these variations into dual-rail fidelity, ideal splitting-ratio-limited Hong–Ou–Mandel visibility, and random-number min-entropy. All realizations satisfy a protocol-performance threshold of 0.95. At 0.99, the tolerated splitting-ratio deviation shrinks from 0.035 for visibility to 0.0035 for min-entropy, and the simulated yields are 68% (T) and 91% (AI1) for coupler visibility and 90% (T) and 66% (AI1) for Y-splitter min-entropy. The tolerance of the target protocol, rather than the device alone, therefore sets the fabrication requirement for silicon spin–photon beam splitters.

## 1. Introduction

Distributed quantum information processing and quantum networks require reliable interfaces between stationary spin qubits and flying photonic qubits. These interfaces enable quantum information to be transferred between distant nodes through single photons.[1-4] Silicon is a promising platform for such interfaces because of its mature semiconductor fabrication technology, compatibility with integrated photonic circuits, and ability to host optically active point defects known as color centers.[2] Several silicon color centers exhibit telecom-band optical transitions and accessible spin states, making them attractive for quantum networking.[2, 5-12] Among them, the T center emits near 1326 nm and offers an optically addressable electron spin.[6] Coherent spin control, single-spin detection, and integration with silicon nanophotonic cavities have been demonstrated.[6, 13-16] More recently, the Al1 center has emerged as another promising telecom-band spin–photon interface. It exhibits optical emission near 1482.44 nm, a spin-1/2 ground state, and higher optical brightness than the T center.[8, 17, 18] These developments provide a foundation for silicon-based quantum nodes. However, connecting multiple nodes requires more than efficient spin–photon interfaces. The emitted photons must also be routed, combined, and interfered with high precision. This creates a need for compact and reliable photonic components that can operate at the emission wavelengths of silicon spin–photon qubits.

Beam splitters are essential components for routing and interfering photons in integrated quantum photonic circuits.[19] They support two-photon interference, photonic state preparation, entanglement generation, and quantum random-number generation. For example, heralded remote entanglement protocols use a balanced beam splitter to interfere photons emitted from separate quantum nodes, followed by single-photon detection.[20-22] Similarly, splitting a single photon between two optical paths enables dual-rail state preparation and path-based quantum random-number generation.[23-25] The performance of these operations depends on the optical power-splitting ratio. Deviations from ideal 50:50 splitting can reduce quantum-state fidelity, Hong–Ou–Mandel (HOM) interference visibility, and the available randomness.[24-26] Therefore, precise and stable splitting is important for reliable quantum photonic operations. However, conventional directional couplers are sensitive to fabrication and environmental variations, which can shift their splitting ratios away from the intended values.[27, 28] Adiabatic designs can improve wavelength tolerance but often require longer coupling regions.[29] These limitations are particularly important for silicon color centers because their emission wavelengths are determined by specific optical transitions. Although limited emitter tuning may be possible,[30] it cannot always compensate for fabrication-induced changes in the beam splitter response. Photonic components must therefore maintain balanced splitting at the target emitter wavelengths despite fabrication imperfections.

Inverse design offers an alternative to conventional photonic design methods by allowing the geometry of a device to be optimized for a desired optical response.[31] Instead of relying on predefined structures, inverse design optimizes the material distribution within a specified region. Gradient-based optimization using the adjoint method efficiently explores large design spaces, while fabrication constraints can be incorporated to improve manufacturability.[32, 33] Previous studies have demonstrated compact, low-loss inverse-designed beam splitters and directional couplers,[34-36] including foundry-fabricated devices[36] and photonic quantum logic gates with measured fidelities.[37] Inverse design has also been applied to photonic structures for solid-state color centers[38, 39] and on-chip nonlinear photon-pair sources.[40] These advances demonstrate the potential of inverse design for integrated quantum photonics. However, many studies focus on optical metrics such as transmission, insertion loss, and splitting ratio. Some also investigate fabrication tolerance using selected geometric variations and examine their effects on quantum-gate fidelity or HOM visibility.[32, 37, 41] Statistical models of fabrication variability are well established for classical silicon photonic circuits,[42, 43] and beam-splitter imperfections have been studied in quantum photonic circuits.[26] Recent work has also linked fabrication noise in conventional and segmented directional couplers to quantum-gate errors.[28, 44] Nevertheless, a systematic connection between fabrication disorder, full-wave optical response, and quantum-protocol yield remains less explored for compact inverse-designed beam splitters operating at silicon color-center wavelengths. Such an analysis is needed to determine whether devices that perform well under nominal conditions can maintain their quantum functionality after fabrication.

Here, we investigate the effects of fabrication disorder on inverse-designed silicon photonic beam splitters and their quantum-protocol performance. We design compact 1×2 Y-splitters and 2×2 directional couplers on a 220 nm silicon-on-insulator (SOI) platform for the T center (1326 nm) and AI1 center (1482.44 nm). We use adjoint-based optimization and three-dimensional finite-difference time-domain (FDTD) simulations to achieve near-balanced splitting with low insertion loss and back-reflection. We then evaluate fabrication robustness using Monte Carlo simulations that include etch bias, line-edge roughness, silicon thickness, and refractive-index variations. The resulting splitting ratios are translated into dual-rail state fidelity, ideal splitting-ratio-limited HOM visibility, and quantum random-number-generator min-entropy. We also calculate simulated manufacturing yield from the fraction of devices meeting specified quantum-performance thresholds. Earlier silicon beam splitters designed for the $C_i$ center at 1452 nm were evaluated using insertion loss, spectral balance, and robustness to dimensional variations.[45] This work provides a statistical assessment of how fabrication disorder affects quantum-protocol performance at emitter-specific wavelengths. We find that mirror symmetry makes the Y-splitter seven to nine times more robust to splitting-ratio variations than the directional coupler, and that the tolerance of the target protocol, which differs by an order of magnitude between HOM visibility and min-entropy, determines how this robustness translates into yield. These findings establish a framework for connecting inverse design, fabrication tolerance, and quantum-protocol yield in silicon spin–photon circuits.

## 2. Results

**Figure 1a** illustrates how photon routing and two-photon interference can be integrated with cavity-coupled silicon color centers and single-photon detectors. A 1×2 Y-splitter divides a photon between two optical paths, while a 2×2 directional coupler combines photons from separate inputs for two-photon interference. Both components must operate at the optical transition wavelengths of the emitters. They must also maintain their intended optical response despite fabrication variations. We address these requirements using adjoint-based inverse design, followed by a statistical analysis of fabrication disorder and quantum-protocol performance.

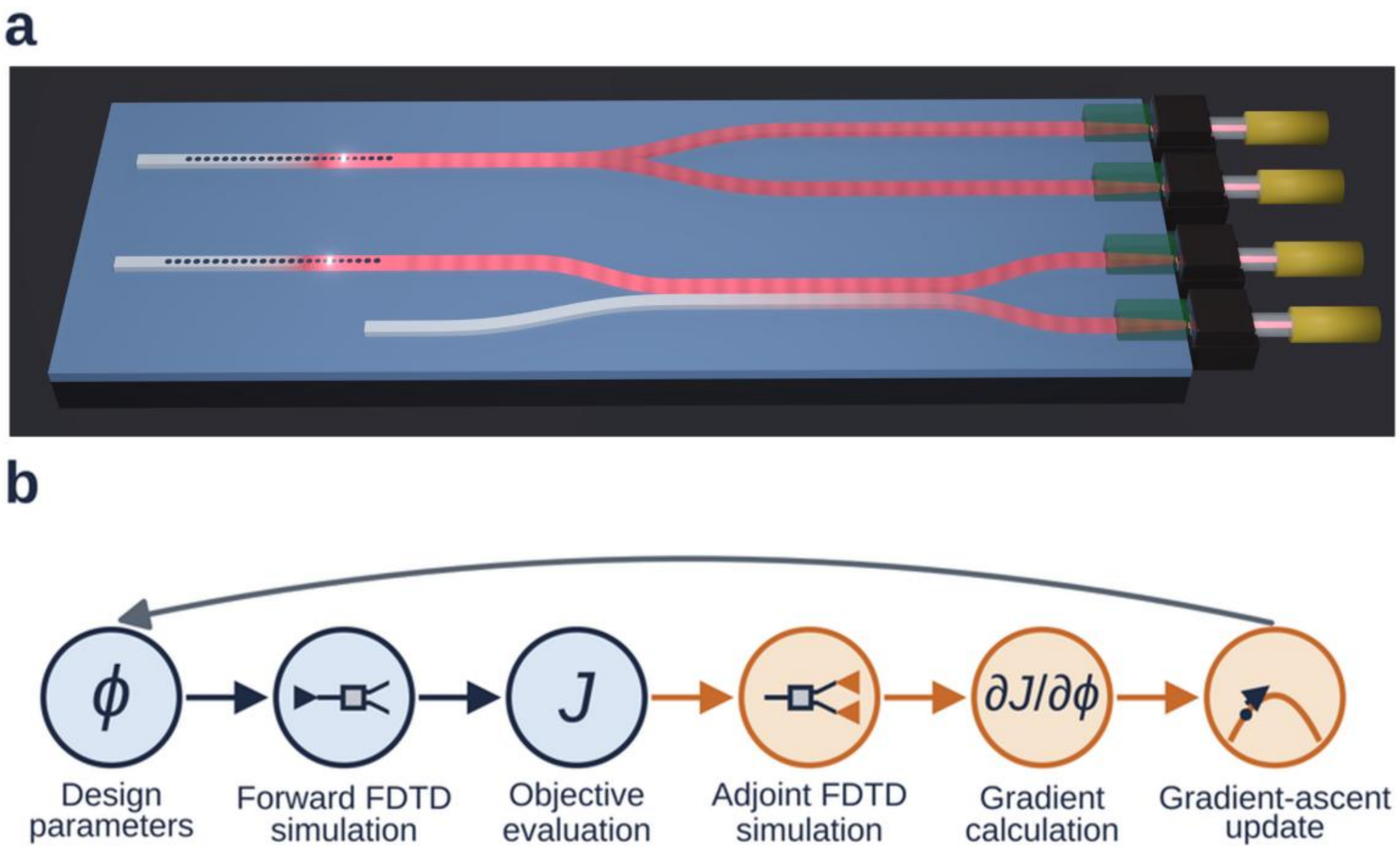

**Figure 1. Conceptual silicon spin–photon quantum photonic circuits and inverse-design workflow.** (a) Schematic of silicon spin–photon interfaces coupled to nanobeam cavities and integrated photonic components for single-photon routing and detection. The upper circuit uses a 1×2 Y-splitter to divide cavity-emitted photons into two output paths. The lower circuit routes the cavity-emitted photons into one input of a 2×2 directional coupler, which enables two-photon interference; the second input is left unexcited for clarity and, in a remote-entanglement node, would receive photons from a second emitter. The outputs are connected to fiber-coupled single-photon detectors through edge couplers. (b) Adjoint-based inverse-design workflow. The optical response is calculated using forward finite-difference time-domain (FDTD) simulations and evaluated against the objective function $J$. An adjoint simulation provides the gradient $\partial J/\partial \phi$ with respect to the level-set function $\phi$, which is used to update the design iteratively; the update ascends the optimization objective $-J$ (Section S1).

## 2.1. Inverse-designed photonic components and nominal optical performance

We first investigate whether inverse design can provide compact beam splitters with high transmission and accurate power splitting at the emission wavelengths of silicon spin–photon qubits. The devices are designed on a silicon-on-insulator (SOI) platform with a 220 nm silicon device layer. We target the T center at 1326 nm and the AI1 center at 1482.44 nm. Rather than relying on conventional waveguide geometries, the adjoint-based optimization iteratively modifies the silicon–air distribution to achieve the desired optical response. Each iteration uses forward and adjoint finite-difference time-domain (FDTD) simulations to evaluate the objective function and update the design parameters (**Figure 1b**).

The resulting structures exhibit distinct geometries that reflect their optical functions. The Y-splitter forms a mirror-symmetric branching region, while the directional coupler consists of two closely spaced waveguides with periodically varying widths (**Figures 2a,b**). Their optimized design regions are 1.7 × 1.7 μm² and 13 × 2 μm², corresponding to footprints of approximately 2.9 and 26 μm², respectively. These dimensions exclude the access waveguides. The simulated optical intensity distributions confirm that the Y-splitter divides the incident field between two output arms, while the directional coupler transfers optical power between its bar and cross ports (**Figures 2c,d**). The optimization objectives increase rapidly during the initial iterations, followed by gradual improvements toward the maximum iteration counts of 150 for the Y-splitter and 80 for the directional coupler (Figure S1, Supporting Information).

Compactness alone is not sufficient for quantum photonic applications. The devices must also transmit photons efficiently while maintaining the intended splitting ratio. We therefore evaluate the total transmission efficiency, η, as the fraction of input power collected in the fundamental transverse-electric ($TE_0$) modes of the output waveguides. The corresponding excess insertion loss is $IL = -10 \log_{10}(\eta)$. We define the splitting ratio, r, as the fraction of transmitted power directed into one output arm of the Y-splitter or the cross port of the directional coupler. Balanced operation corresponds to r = 0.5.

At the emitter wavelengths, both device types exhibit high transmission with low excess insertion loss (**Figures 2e,f**). The T-center and Al1-center Y-splitters transmit 93.9% and 96.8% of the input power, corresponding to excess losses of 0.275 and 0.143 dB, respectively. Both maintain a nominal splitting ratio of 0.500 through their mirror-symmetric geometry, so their transmission is evaluated as twice the $TE_0$ power in one output arm (Section S2, Supporting Information). The directional couplers achieve transmission efficiencies of 90.3% and 95.1%, with corresponding excess losses of 0.442 and 0.218 dB. However, their splitting ratios differ. The Al1-center coupler is nearly balanced at r = 0.501, while the T-center coupler exhibits a small nominal imbalance at r = 0.470 (Table S1, Supporting Information). The devices also maintain low back-reflection, ranging from $1.0 \times 10^{-5}$ to $4.5 \times 10^{-4}$ of the incident power. For the directional couplers, leakage into the first higher-order guided mode ($TE_1$) remains approximately $10^{-3}$ or lower across the investigated spectral range. These results indicate that most transmitted power remains in the desired fundamental waveguide modes (Figure S2, Supporting Information).

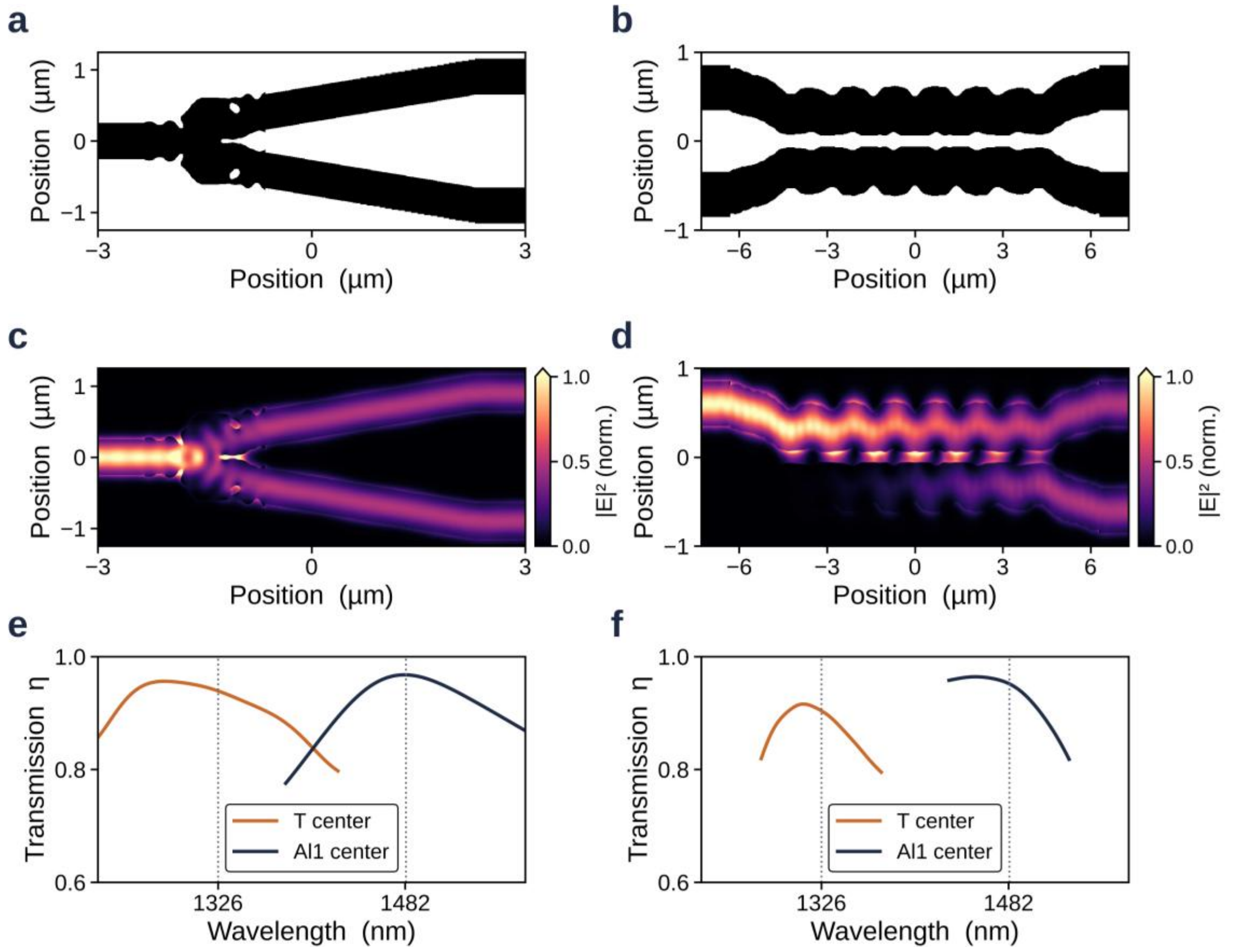

**Figure 2. Optimized geometries and optical performance of the inverse-designed beam splitters.** (a,b) Optimized binary silicon–air layouts of the Al1-center (a) 1×2 Y-splitter and (b) 2×2 directional coupler; black and white regions represent silicon and air, respectively. (c,d) Simulated steady-state intensity distributions, $|E|^2$, at 1482.44 nm in the (c) Y-splitter and (d) directional coupler, normalized to the peak intensity of the input mode; the color scale saturates at the field-enhancement spots at the junction tip and in the coupling gap. The vertical scale of (b) and (d) is expanded about three times relative to the horizontal scale. (e,f) Total transmission η versus wavelength for the (e) Y-splitters and (f) directional couplers optimized at 1326 nm (T center, orange) and 1482.44 nm (Al1 center, navy). Vertical dotted lines indicate the operating wavelengths.

Although the devices are optimized at specific emitter wavelengths, their maximum transmission does not always occur exactly at those wavelengths. This difference is most pronounced for the T-center Y-splitter, whose transmission peaks near 1280 nm. Nevertheless, both Y-splitters maintain excess insertion loss below 0.5 dB over broad wavelength ranges of 1238–1376 nm for the T center and 1426–1563 nm for the Al1 center. The directional couplers also maintain near-balanced splitting, with $|r - 0.5| \leq 0.05$ over 1304–1333 nm and 1464–1497 nm, respectively. These operating windows include the corresponding emitter wavelengths and indicate tolerance to moderate spectral deviations.

Together, these results demonstrate that inverse design can produce compact, low-loss photonic components tailored to silicon spin–photon interfaces. However, the two geometries achieve their optical functions through different physical mechanisms. The Y-splitter maintains balanced power division through geometric symmetry, while the directional coupler relies on wavelength-dependent coupling between adjacent waveguides. This distinction raises an important question: how reliably can each device preserve its splitting ratio when fabrication imperfections alter its geometry and optical properties?

### 2.2. Splitting-ratio robustness under fabrication disorder

The nominal optical performance establishes that both device geometries can provide efficient photon routing at the target emitter wavelengths. However, fabrication imperfections can alter their optical response and shift the splitting ratio away from the intended 50:50 condition. The impact of these changes depends on how the splitting ratio varies with wavelength and how the device responds to geometric and material perturbations.

The wavelength-dependent responses reveal a clear distinction between the two geometries (**Figures 3a,b**). The Y-splitter maintains a splitting ratio close to 0.5 across the investigated spectral ranges. Its mirror-symmetric geometry directs equal power into the two output arms as long as the symmetry is preserved. In contrast, the directional coupler relies on wavelength-dependent coupling between adjacent waveguides. Near the emitter wavelengths, its splitting ratio changes by approximately 3.1–3.4 × $10^{-3}$ per nanometer. Consequently, the coupler remains within ±0.01 of balanced splitting over only about 5.9 nm for the T-center design and 6.5 nm for the Al1-center design.

This wavelength sensitivity makes the position of the nominal balance point particularly important. For the T-center coupler, balanced splitting occurs near 1317.7 nm, approximately 8 nm below the target emission wavelength. As a result, the splitting ratio is 0.470 at the nearest simulated spectral sample and approximately 0.472 when interpolated to 1326 nm. This offset becomes apparent when the optimized geometry is evaluated using a finer validation grid. On the original optimization grid, the same design is nearly balanced at the target wavelength, with r = 0.496 at 1326 nm (Table S4, Supporting Information). In comparison, the Al1-center coupler exhibits a balance-point shift of less than 0.1 nm between the optimization and validation simulations. Its splitting ratio remains within ±0.035 of balance over 1470–1493 nm, covering the approximately 1479–1483 nm zero-phonon-line range reported for individual Al1 centers.

To determine how these nominal responses change under fabrication disorder, we generate 100 independent Monte Carlo realizations for each device and emitter wavelength. Each realization includes four sources of variation: global etch bias, line-edge roughness, silicon thickness, and refractive index. The corresponding standard deviations are 2 nm for etch bias, 3 nm for silicon thickness, and 0.005 for refractive index. Line-edge roughness is modeled with a root-mean-square amplitude of 3 nm and a correlation length of 50 nm. These parameters represent assumed local fabrication variations, informed by reported silicon-on-insulator roughness statistics[46] and representative silicon photonic process variations.[42, 43] The disorder model and simulation procedures are described further in the Methods and Section S3 of the Supporting Information.

The resulting splitting-ratio distributions show that the Y-splitter is substantially more robust than the directional coupler (**Figure 3c**). The T-center and Al1-center Y-splitters exhibit standard deviations of 0.0023 and 0.0034, respectively. In comparison, the corresponding directional couplers show standard deviations of 0.020 and 0.024. Thus, the Y-splitter exhibits approximately seven- to ninefold smaller splitting-ratio fluctuations under the investigated fabrication conditions (Table S2 and Figure S3, Supporting Information).

The physical origin of this difference becomes clearer when the disorder sources are examined separately. Uniform variations in etch bias, silicon thickness, and refractive index preserve the symmetry between the two Y-splitter arms. These perturbations can modify the overall transmission but have little effect on the relative output powers. The individual splitting-ratio shifts associated with the four scalar disorder parameters remain below 0.001. A multiple linear regression using these parameters explains only approximately 4–6% of the Y-splitter variance and is not statistically significant (F-test p = 0.25 and 0.40). The remaining fluctuations are consistent with local symmetry-breaking perturbations, particularly the spatial pattern of line-edge roughness, which cannot be fully represented by a single scalar roughness amplitude.

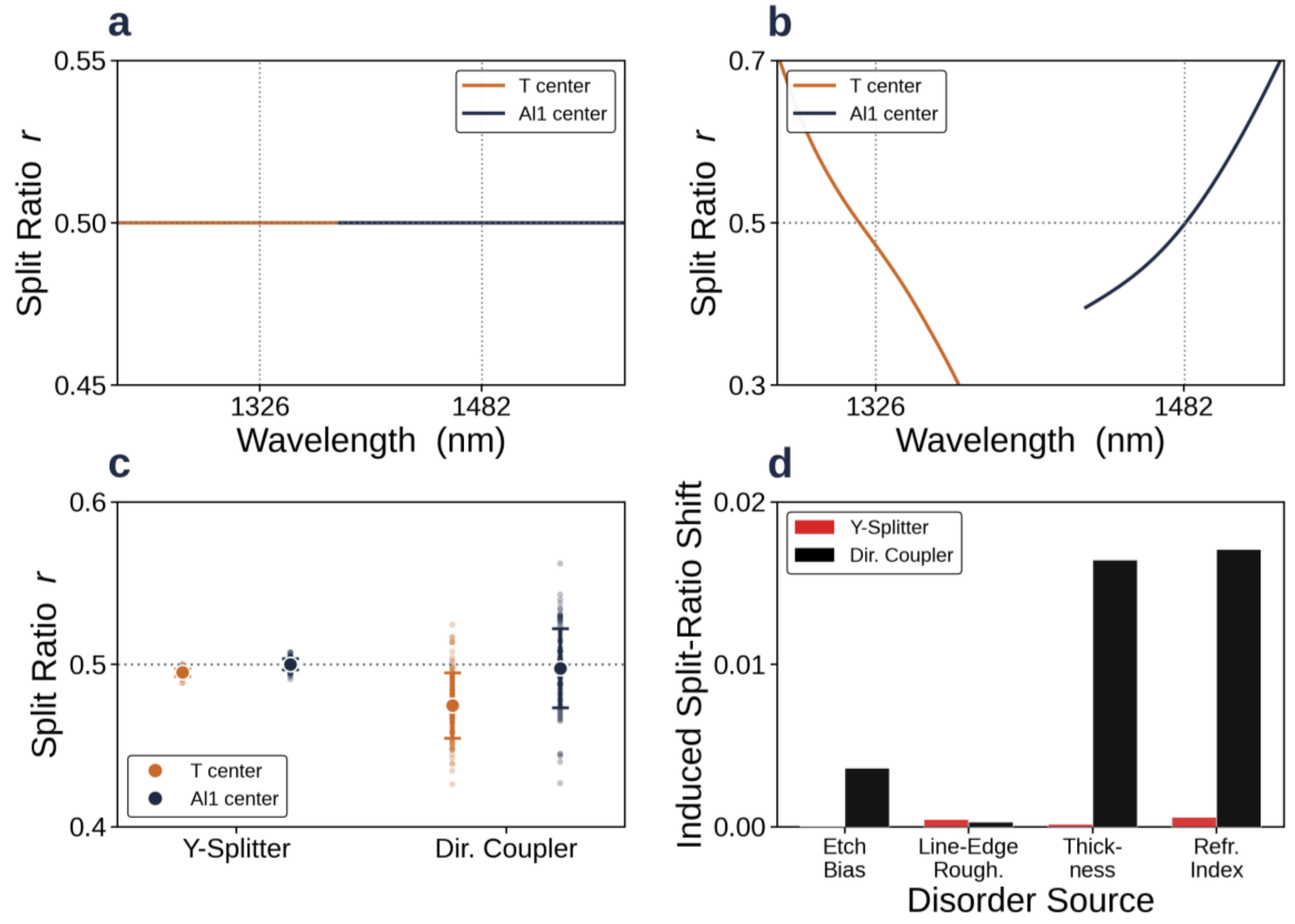

**Figure 3. Wavelength dependence and fabrication robustness of inverse-designed beam splitters.** (a,b) Simulated split ratio as a function of wavelength for the (a) Y-splitter and (b) directional coupler optimized for the T center (orange) and Al1 center (navy). Vertical dotted lines indicate the operating wavelengths, and the horizontal dotted line represents ideal 50:50 splitting. (c) Split-ratio distributions obtained from 100 Monte Carlo realizations for each device and operating wavelength; the T-center Y-splitter distribution is centered at $r$ = 0.495 by a numerical offset of the simulation (Section 2.3). Large markers and error bars show the mean and standard deviation, while smaller markers represent individual realizations. (d) Split-ratio shifts associated with one-standard-deviation variations in etch bias, line-edge roughness, silicon thickness, and refractive index for the Al1-center designs; T-center values are given in Table S3.

The directional coupler responds differently because its splitting ratio depends on the coupling conditions between adjacent waveguides. Variations in geometry and refractive index change the effective modal indices and coupling strength, redistributing optical power between the bar and cross ports. Accordingly, the regression model explains approximately 94% of the splitting-ratio variance for the T-center coupler and 98% for the Al1-center coupler. At the T-center wavelength, refractive-index variation makes the largest contribution, followed by etch bias and silicon thickness. Their squared standardized regression coefficients are 0.52, 0.27, and 0.18, respectively. For the Al1-center coupler, refractive index and silicon thickness contribute nearly equally, with corresponding values of 0.49 and 0.46. Because the sampled disorder variables are weakly correlated, these squared coefficients do not sum exactly to the explained variance. Line-edge roughness contributes comparatively little in both cases (**Figure 3d** and Table S3, Supporting Information).

These results show that the Y-splitter benefits from symmetry-protected power division, while the directional coupler remains sensitive to changes in its optical coupling conditions. However, a small splitting-ratio spread does not necessarily guarantee reliable quantum operation. The acceptable imbalance depends on the quantum protocol and its required performance threshold. We therefore examine how the simulated fabrication-induced variations affect quantum-state fidelity, two-photon interference, and random-number generation.

## 2.3. From fabrication disorder to quantum-protocol performance and yield

Different quantum protocols respond differently to the same splitting imbalance. We translate the simulated splitting ratios into three quantum-protocol figures of merit: dual-rail state fidelity, Hong–Ou–Mandel (HOM) visibility, and quantum random-number-generator min-entropy. We assume ideal photon indistinguishability and detection, while treating optical loss separately (Section S4, Supporting Information). The Y-splitter is evaluated for single-photon path encoding and random-number generation, whereas the directional coupler is evaluated primarily for two-photon interference.

When a single photon is divided between two optical paths, it forms a dual-rail state $\sqrt{r}|1,0\rangle + \sqrt{(1-r)}|0,1\rangle$. After compensating for the relative phase between the paths, the fidelity with an ideal balanced state $(|1,0\rangle + |0,1\rangle)/\sqrt{2}$ is[24]

$$F(r) = \frac{1}{2} + \sqrt{r(1-r)}. \qquad (1)$$

For two indistinguishable photons entering separate inputs of an ideal coupler, HOM interference[47] yields the split-ratio-limited visibility[48]

$$V(r) = \frac{2r(1-r)}{r^2 + (1-r)^2}. \qquad (2)$$

This expression assumes the phase relationship of an ideal symmetric beam splitter. For the nominal directional couplers, the simulated bar–cross phase differences depart from the ideal quadrature condition by only 1.5° for the T-center design and 1.0° for the Al1-center design. These deviations reduce the calculated visibility by less than 0.2% (Section S4, Supporting Information). In practice, photon distinguishability and deviations in the full complex scattering response can further reduce the observed visibility.[41, 49, 50]

For path-based quantum random-number generation, the splitting ratio determines the probability of detecting a photon in either output path.[25] The corresponding min-entropy per detection event is

$$H_{\min}(r) = -\log_2[\max(r,\, 1-r)]. \qquad (3)$$

A balanced splitter produces one bit of min-entropy per detection event under ideal detection conditions. Any splitting imbalance reduces this value and increases the amount of randomness extraction needed to obtain unbiased output bits.[51]

Although all three figures of merit reach their maximum values at r = 0.5, they respond differently to deviations from balance (**Figure 4a**). Dual-rail fidelity and ideal HOM visibility decrease quadratically near the balanced condition, while min-entropy decreases approximately linearly with the absolute splitting error. For example, a deviation of $|r - 0.5| = 0.0176$ reduces min-entropy to 0.95 bit, whereas HOM visibility reaches the same threshold at a deviation of approximately 0.080. At a stricter requirement of 0.99, the allowable deviations are 0.0995 for dual-rail fidelity, 0.035 for HOM visibility, and only 0.0035 for min-entropy (Section S4, Supporting Information). These differences arise from the quantum-protocol requirements themselves, rather than from the particular device geometry.

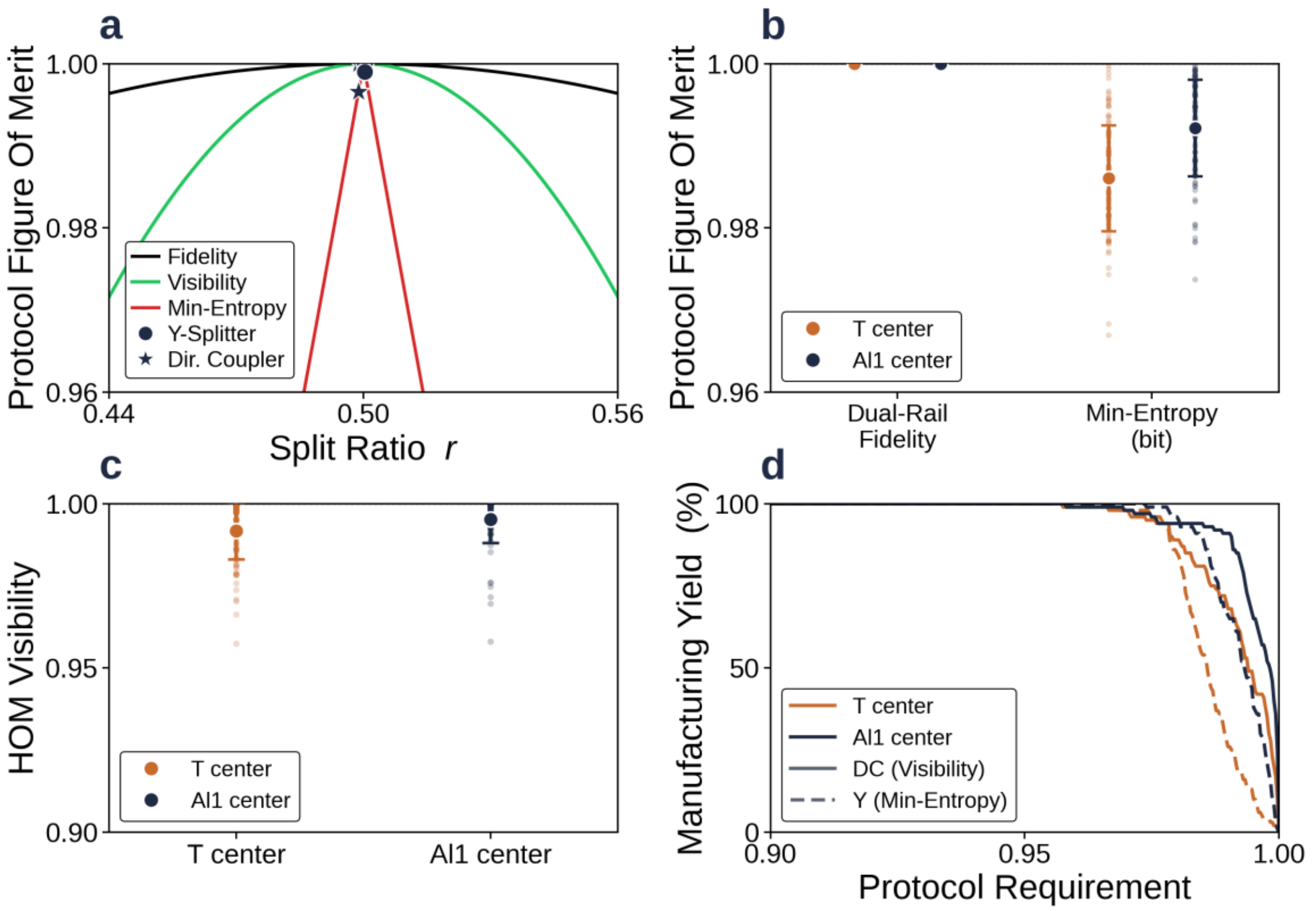

**Figure 4. Quantum-protocol performance and simulated yield under fabrication disorder.** (a) Protocol figures of merit versus split ratio *r*: dual-rail state fidelity *F* (black), HOM visibility *V* (green), and random-number min-entropy $H_{min}$ (red), each maximized at $r = 0.5$. Markers indicate the nominal Monte Carlo split ratios of the Al1-center Y-splitter (circle) and directional coupler (star), plotted on the min-entropy curve. (b) Dual-rail fidelity and min-entropy of the Y-splitter across the 100-realization ensemble for the T-center (orange) and Al1-center (navy) designs. (c) HOM visibility of the T-center and Al1-center directional couplers across the same ensemble. In (b) and (c), large markers and error bars denote the mean and ±1 standard deviation, upper bars that would exceed the physical bound of 1 being clipped by the axis for these skewed distributions, and small markers show individual realizations. (d) Simulated yield, the fraction of realizations meeting a given protocol requirement, versus that requirement: directional-coupler (DC) visibility (solid) and Y-splitter min-entropy (dashed) for the T center (orange) and Al1 center (navy). The T-center Y-splitter data in (b) and (d) are shown for the raw ensemble, which carries the numerical offset discussed in Section 2.3.

Applying these relations to the Monte Carlo ensembles reveals how fabrication disorder affects each quantum operation (**Figures 4b,c**). Dual-rail fidelity remains above 0.9998 for every Y-splitter realization and above 0.994 for every directional-coupler realization. The ideal splitting-ratio-limited HOM visibility of the directional couplers remains above 0.957. The Y-splitters also maintain high min-entropy, with minimum values of 0.967 bit for the T-center design (0.981 bit when referenced to its balanced nominal, as discussed below) and 0.974 bit for the Al1-center design (Table S2, Supporting Information). These results indicate that the devices maintain high quantum-protocol performance under the assumed fabrication conditions, although their sensitivity to splitting imbalance differs.

To evaluate this difference more directly, we define simulated manufacturing yield as the fraction of Monte Carlo realizations that meet a specified quantum-protocol requirement. At a threshold of 0.95, all 100 realizations of each design satisfy the requirements of their intended applications. However, stricter thresholds reveal substantial differences in yield (**Figure 4d**). At a HOM-visibility requirement of 0.99, the T-center directional coupler achieves a yield of 68%, compared with 91% for the Al1-center coupler. The corresponding 95% confidence intervals are 58–77% and 84–96%, respectively. Applying the nominal coupler phase to all realizations changes these values only to 64% and 89% (Sections S4 and S7), and the heralded Barrett–Kok fidelity itself does not depend on this phase, which only adds a correctable phase to the heralded state (Section S4). For the Y-splitters, the min-entropy yields at 0.99 bit are 26% for the raw T-center ensemble and 66% for the Al1 center; the T-center value is limited by a numerical offset of the simulation and rises to 90% when referenced to the balanced nominal of the symmetric design, as discussed below. In contrast, dual-rail fidelity remains above 0.99 for every simulated realization. The complete yield curves and statistical confidence intervals are provided in Figure S4 and Table S5 of the Supporting Information.

These visibility requirements are relevant to heralded entanglement between remote quantum nodes. In the idealized Barrett–Kok protocol, splitting imbalance can reduce the fidelity of the heralded Bell state, depending on the detector outcomes. For opposite-detector heralds, the fidelity is $[1 + V(r)]/2$, while same-detector heralds retain ideal fidelity under the assumed conditions (Section S4, Supporting Information). Thus, a splitting-ratio-limited visibility of 0.99 corresponds to a Bell-state fidelity of at least 0.995 within this model. Actual entanglement fidelity also depends on photon indistinguishability and other imperfections, which are not included in the present yield estimates.[21, 22]

The different directional-coupler yields reveal an important role of nominal splitting accuracy. Although the Al1-center coupler has a slightly larger splitting-ratio standard deviation (0.024) than the T-center coupler (0.020), it achieves a higher visibility yield. Its ensemble mean is close to balance at $r = 0.498$, while the T-center coupler is shifted to $r = 0.475$. This offset arises from the displacement of its nominal balance point relative to the emitter wavelength, as discussed in Section 2.2. If the T-center distribution were recentered at r = 0.5 while retaining its spread, the estimated visibility yield would increase from 68% to approximately 90%. The lower yield therefore results primarily from nominal imbalance rather than greater fabrication sensitivity.

The Y-splitter min-entropy results show the same sensitivity to the operating point, but here the offset is numerical rather than physical. The T-center Y-splitter has the narrowest splitting-ratio distribution, yet its simulated ensemble is centered at r = 0.495 instead of 0.5. Its geometry is mirror-symmetric by construction, so the designed device is balanced. The offset arises from the discretization of the design-region permittivity on the FDTD grid: at a few grid nodes located exactly midway between two permittivity samples, the nearest-sample interpolation selects samples that are not mirror images of each other, which slightly breaks the symmetry of the simulated structure (Section S3, Supporting Information). Because min-entropy is highly sensitive to small deviations near r = 0.5, this numerical offset lowers the 0.99-bit yield of the raw ensemble to 26%. Referenced to the balanced nominal, the T-center Y-splitter reaches a min-entropy yield of 90% (95% confidence interval 82–95%), which reflects its physical fabrication tolerance.

Together, these findings show that quantum-protocol yield depends on the spread of fabrication-induced splitting errors relative to the tolerance window of the target protocol, and on the position of the nominal operating point within that window. A device can be highly robust to disorder but still have a low yield if its splitting ratio is not centered in a narrow window. Conversely, a nominally balanced device may lose yield when fabrication variations are large. Inverse-designed quantum photonic components should therefore be evaluated using application-specific performance requirements rather than nominal optical metrics alone.

## 3. Discussion

### 3.1. Comparison with existing beam splitters

Table 1 compares the nominal optical performance of our devices with previously reported silicon photonic beam splitters and couplers. The Y-splitter achieves simulated excess insertion losses of 0.14–0.27 dB within a compact design region of 2.9 μm². Its footprint is smaller than that of a previously reported inverse-designed 1×2 divider (6.8 μm²), which exhibits a simulated loss of 0.33 dB,[34] and comparable to that of a parameterized Y-junction (3.2 μm²).[52] A previously reported Y-junction also demonstrated an average measured insertion loss of 0.28 ± 0.02 dB across an 8-inch wafer,[53] providing a useful experimental benchmark for the simulated losses obtained here.

The directional coupler achieves simulated excess losses of 0.22–0.44 dB within a 26 μm² design region. This footprint is smaller than that of a previously reported 100 μm² inverse-designed 2×2 coupler,[35] but larger than other compact inverse-designed splitters with footprints of 3.6 and 1.69 μm².[36, 37] The present coupler maintains a splitting ratio within ±0.05 of balance over wavelength ranges of approximately 28–33 nm. For comparison, an adjoint-optimized adiabatic coupler uses a 30 μm coupling region to achieve a simulated 204 nm operating bandwidth within 3 ± 0.5 dB.[29] These comparisons indicate that the present devices offer competitive nominal optical performance, although differences in operating wavelength, design constraints, and evaluation methods prevent a direct ranking of their performance.

A key distinction of this work is its focus on the emission wavelengths of silicon spin–photon centers. Most devices in Table 1 were designed for operation near 1520–1550 nm, where the input wavelength can often be adjusted to compensate for small shifts in the splitting balance. In contrast, emitter-based quantum photonic circuits must operate near the characteristic optical transitions of the T and Al1 centers at 1326 and 1482.44 nm. Accurate splitting at these wavelengths is therefore particularly important, since the photon wavelength cannot always be freely adjusted to match the balance point of a fabricated device.

Beyond conventional optical performance, previous studies have demonstrated inverse-designed quantum photonic gates with measured fidelities and robustness to etch variations,[37] as well as a spatial-mode beam splitter with measured Hong–Ou–Mandel (HOM) visibility exceeding 99%.[41] Other studies have used Monte Carlo simulations to investigate fabrication-induced quantum-gate errors in conventional and segmented directional couplers, including designs optimized for gate fidelity under waveguide-width variations.[28, 44] The present work connects these approaches by evaluating inverse-designed silicon beam splitters at specific emitter wavelengths and translating multiple sources of fabrication disorder into statistical quantum-protocol yields. This provides an additional measure of device reliability beyond the nominal optical benchmarks summarized in Table 1.

**Table 1. Comparison of the optical performance of inverse-designed and conventional silicon photonic beam splitters and couplers.** The devices developed in this work are shown in bold. Losses are simulated unless marked "meas.", and footprints refer to the device design regions, excluding access waveguides; multimode-interference couplers are not included. For the 1×2 divider,[34] the loss is the simulated average over 1530–1570 nm (0.44 dB measured) and r = 0.50 holds by symmetry; the 2×2 coupler values[35] are from 2.5D simulations, for the top and bottom inputs. The coupler values of this work are given at the spectral samples closest to the target wavelengths (Table S1).

| Device (this work in bold) | Method | λ (nm) | Footprint (μm²) | Excess IL (dB) | Split ratio r |
|---|---|---|---|---|---|
| **Y-splitter, T center** | inverse design | 1326 | 1.7×1.7 (2.9) | 0.27 | 0.500 |
| **Y-splitter, Al1 center** | inverse design | 1482.44 | 1.7×1.7 (2.9) | 0.14 | 0.500 |
| 1×2 divider[34] | inverse design | 1550 | 2.6×2.6 (6.8) | 0.33 | 0.50 |
| Y-junction[52] | parameterized | C band | 1.4×2.3 (3.2) | <0.36 | arbitrary (layout-set) |
| **Dir. coupler, T center** | inverse design | 1326 | 13×2 (26) | 0.44 | 0.470 |
| **Dir. coupler, Al1 center** | inverse design | 1482.44 | 13×2 (26) | 0.22 | 0.501 |
| Dir. coupler[32] | inverse design | 1520 | not reported | 0.46 | 0.5 (target) |
| 2×2 coupler[35] | inverse design (2.5D) | 1550 | 10×10 (100) | 0.20 / 0.32 | 0.51 / 0.49 |
| Dir. coupler[36] | inverse design | 1545 (design); ≈1505 (meas.) | 3.0×1.2 (3.6) | 0.5 (meas.) | imbalance <10% (meas.) |
| 2×2 splitter[37] | inverse design | 1550 | 1.3×1.3 (1.69) | not reported | 0.5 (target) |

### 3.2. Design implications for emitter-based circuits

Our results show that the splitting accuracy required for reliable quantum operation depends strongly on the intended protocol. Although dual-rail fidelity, HOM visibility, and random-number min-entropy all reach their ideal values at a balanced splitting ratio, they tolerate different levels of imbalance. At a performance requirement of 0.99, min-entropy allows a splitting-ratio deviation of only 0.0035, approximately ten times smaller than the tolerance for HOM visibility (Section S4, Supporting Information). Thus, achieving nominal 50:50 splitting is only the first requirement. The device must also preserve the necessary balance under fabrication variations, with a tolerance determined by the quantum protocol it is intended to support.

The comparison between the Y-splitter and directional coupler further demonstrates the importance of device geometry. The mirror-symmetric Y-splitter maintains nearly equal power division under perturbations that preserve symmetry between its output arms. As a result, its splitting ratio is relatively insensitive to uniform changes in etch bias, silicon thickness, and refractive index. The directional coupler, in contrast, relies on evanescent coupling between adjacent waveguides. Changes in geometry or refractive index modify the coupling conditions and can shift the splitting ratio away from balance. These differences suggest that geometric symmetry offers a useful route to fabrication robustness when the optical function allows it. For directional couplers, robustness must be incorporated more explicitly into the design. Existing robust inverse-design methods that evaluate perturbed geometries, including over- and under-etched structures,[33] could be extended to account for thickness and refractive-index variations. Broadband optimization may also reduce sensitivity to fabrication-induced spectral shifts.[36]

However, fabrication robustness alone does not guarantee high quantum-protocol yield. The T-center directional coupler exhibits a smaller splitting-ratio spread than the Al1-center coupler but achieves a lower HOM-visibility yield because its nominal balance point is displaced from the emitter wavelength (Section 2.2). Nominal splitting accuracy and fabrication-induced variation must therefore be considered together. Post-fabrication tuning using reconfigurable photonic circuits offers one way to correct a residual imbalance,[54] although it introduces additional components and control requirements.

These observations suggest that future inverse-design approaches should consider quantum-protocol performance directly during optimization. Instead of optimizing only nominal transmission and splitting ratio, the objective function could incorporate fabrication perturbations and the expected performance of the intended quantum operation. For example, a directional coupler could be optimized to maintain high HOM visibility across an ensemble of geometric and material variations. Related strategies have been explored for segmented couplers optimized for quantum-gate fidelity under fabrication errors.[44] Extending such approaches to full-wave inverse-designed structures could improve the fraction of devices that satisfy a specified quantum-performance requirement.

Optical loss introduces another important consideration. In the ideal conditional models used here, balanced loss reduces the probability of successful photon transmission and detection without directly reducing the normalized conditional fidelity. For protocols requiring the detection of two photons,[20, 21] the transmission contribution to the success probability scales approximately as $\eta^2$, corresponding to 0.82 and 0.90 for the T-center and AI1-center directional couplers, respectively (Section S5, Supporting Information). Therefore, a device with high HOM visibility may still limit the rate of successful entanglement generation if its transmission is low. Future optimization should consider both quantum-state quality and photon transmission rather than treating them as independent design objectives.

### 3.3. Scope and limitations

The present analysis isolates the effect of splitting-ratio variations on quantum-protocol performance. In practice, HOM interference also depends on the complex scattering amplitudes of the beam splitter and the indistinguishability of the interfering photons. Although the nominal directional couplers exhibit only small deviations from the ideal phase relation (Section S4, Supporting Information), fabrication-induced asymmetry may introduce additional phase errors. Differences in photon frequency, polarization, and temporal wavepacket can further reduce the observed visibility. A more complete evaluation would therefore require the full complex scattering matrix for each perturbed device, together with the spectral and temporal properties of the emitted photons. This is particularly important when photons from independent silicon spin–photon centers are interfered for remote entanglement.

The simulated yields also depend on the assumed fabrication-disorder model. Our Monte Carlo analysis includes local variations in etch bias, line-edge roughness, silicon thickness, and refractive index, but does not capture every source of fabrication uncertainty. Systematic process offsets, wafer-scale variations, and correlations between neighboring devices may introduce additional changes in optical performance. Previous experimental studies have reported substantial spectral shifts in fabricated inverse-designed couplers,[36] illustrating the importance of these effects. Moreover, the 100-realization ensembles introduce statistical uncertainty into the estimated yields, as quantified in Section S7 of the Supporting Information. The reported values should therefore be interpreted as simulated yields under the specified disorder assumptions rather than predictions of actual manufacturing yield.

Finally, the devices are evaluated individually using fundamental waveguide-mode excitation. Their performance in complete silicon quantum photonic circuits will also depend on coupling to emitter cavities, waveguide transitions, propagation losses, and single-photon detection. Future work should combine fabrication-robust inverse design with full scattering-matrix simulations, integrated emitter–cavity–waveguide modeling, and experimental fabrication and characterization. These steps will be necessary to establish whether the predicted device-level robustness translates into reliable quantum operation in integrated silicon spin–photon systems.

## 4. Conclusions

We have investigated the optical performance and fabrication robustness of inverse-designed silicon photonic beam splitters tailored to the emission wavelengths of T and Al1 spin–photon centers. The optimized Y-splitters and directional couplers achieve compact footprints and low excess insertion losses of 0.14–0.44 dB. Using Monte Carlo simulations that account for variations in etch bias, line-edge roughness, silicon thickness, and refractive index, we find that the mirror-symmetric Y-splitter exhibits seven- to ninefold smaller splitting-ratio fluctuations than the directional coupler. By translating these variations into quantum-protocol figures of merit, we show that fabrication robustness alone does not determine device yield: the tolerance of the target protocol, ten times stricter for min-entropy than for HOM visibility at a requirement of 0.99, sets how much splitting error a device can afford. At this requirement, the directional couplers achieve simulated HOM-visibility yields of 68% and 91% for the T and Al1 centers, respectively, and the Y-splitters achieve min-entropy yields of 90% and 66%, the T-center value being referenced to the balanced nominal of the symmetric design. The lower yield of the T-center coupler reflects the displacement of its nominal balance point from the emitter wavelength rather than a larger sensitivity to disorder. Our findings establish a numerical framework for connecting inverse-designed photonic components and fabrication disorder to quantum-protocol yield, providing a basis for developing more reliable silicon spin–photon quantum photonic circuits.

**Methods:**

**Inverse design.** The photonic components were designed using the Tidy3D three-dimensional FDTD solver (Flexcompute Inc.) with adjoint-based gradient optimization. Each device was optimized within a rectangular design region on an SOI platform consisting of a 220 nm silicon device layer ($n_{Si}$ = 3.46) on a 2 μm silica buried-oxide layer ($n_{SiO2}$ = 1.444), with air as the upper cladding and in the etched regions. The input and output waveguides were 500 nm wide and supported the fundamental transverse-electric ($TE_0$) mode. A level-set parametrization was used to describe the silicon–air distribution within the design region, with the permittivity represented on a 4 nm grid for the Y-splitter and on a 6.7 nm (T center) or 7.5 nm (Al1 center) grid for the directional coupler.[32] The optimization objective for the Y-splitter was to maximize the total $TE_0$ transmission into the two output waveguides while minimizing back-reflection. For the directional coupler, the objective penalized the deviations of the bar and cross powers from 0.5, the excess loss, and the residual conversion into the first-order mode; the explicit objective functions are given in Section S1 of the Supporting Information. Minimum feature sizes were penalized through level-set gap and curvature penalties,[55] with targets of 140 nm (Y-splitter) and 50 nm (directional coupler); because these penalties are soft, the final Y-splitter layouts contain isolated air features down to about 60 nm, and both devices are compatible with electron-beam lithography. The devices were optimized separately at 1326 nm and 1482.44 nm, the zero-phonon lines of the T center and of an isolated Al1 center,[8] respectively. The optimization was performed for up to 150 iterations for the Y-splitter and 80 iterations for the directional coupler. Additional details of the optimization procedure are provided in Sections S1 and S6 and Table S4 of the Supporting Information.

**High-resolution electromagnetic validation.** Following optimization, each device was simulated again on a finer Yee grid. The maximum cell size was $\lambda_c/(74\ n_{Si})$ for the Y-splitter (about 5.2 and 5.8 nm at 1326 and 1482.44 nm) and $\lambda_c/(57\ n_{Si})$ for the directional coupler (about 6.7 and 7.5 nm), where $\lambda_c$ is the operating wavelength. The nominal simulations were performed with run times of 1.5 ps (Y-splitter) and 5.0 ps (directional coupler), and the spectra were sampled over ±100 nm (Y-splitter, 201 points) and ±50 nm (directional coupler, 41 points) around the operating wavelengths; the nominal metrics are reported at the spectral samples closest to the operating wavelengths, within 1 nm of them (Table S1). The optical response was obtained from the power coupled into the fundamental $TE_0$ modes of the output waveguides. The total transmission efficiency was calculated by summing the output-mode powers and normalizing them to the input power; for the mirror-symmetric Y-splitter, the validation simulation monitored one output arm, and the total transmission was taken as twice its power. The split ratio was obtained from the fraction of transmitted power directed into the selected output port. Back-reflection was evaluated from the optical power coupled into the backward-propagating fundamental mode at the input. For the directional coupler, each input was excited in a separate simulation, and the complex $TE_0$ amplitudes of the bar and cross outputs were used to evaluate their relative phase (Section S4). Additional validation results are provided in Section S2 of the Supporting Information.

**Fabrication-tolerance analysis.** The effects of fabrication variations were evaluated using Monte Carlo simulations. For each device and operating wavelength, 100 independent disorder realizations were generated using the same fixed random seed for all four ensembles. Each realization included four fabrication variations: global etch bias, line-edge roughness (LER), silicon thickness, and refractive index. The global etch bias was modeled with a standard deviation of 2 nm. The LER was represented by a spatially correlated random field with a root-mean-square amplitude of 3 nm and a correlation length of 50 nm. Silicon thickness and refractive index were varied with standard deviations of 3 nm and 0.005, respectively. Etch bias and LER were introduced through a uniform boundary displacement and a spatially correlated perturbation of the level-set function, respectively. Each perturbed geometry was then evaluated with three-dimensional FDTD simulations on the same spatial grid as the corresponding high-resolution validation, with a run time of 1.0 ps, and each ensemble was referenced to a nominal simulation computed with identical settings. The resulting optical responses were used to determine the split-ratio distributions and evaluate device sensitivity to fabrication disorder.

To quantify the contribution of each fabrication parameter, we performed multiple linear regression of the split ratio against the four disorder variables. The regression coefficients were standardized to compare the relative influence of each source, and the significance of each regression was assessed with an F-test. The resulting split-ratio distributions were also used to calculate quantum-protocol figures of merit and simulated yield. For a given performance threshold, the yield was calculated as the fraction of realizations satisfying that requirement. Statistical confidence intervals were obtained using the exact Clopper–Pearson binomial method.[56] Further details of the disorder model, numerical simulations, and yield analysis are provided in Sections S3, S6, and S7 of the Supporting Information.

***Supporting Information.*** Inverse-design framework and optimization convergence (S1); high-resolution electromagnetic validation and nominal device performance (S2); fabrication-tolerance analysis and variance budget (S3); quantum-protocol figures of merit, analytical extensions, and simulated yield (S4); back-reflection, optical loss, and integration considerations (S5); extended computational methods and simulation parameters (S6); statistical uncertainty of Monte Carlo yield estimates (S7).

**Data Availability.** The data that support the findings of this study are available from the corresponding author upon reasonable request.

**Code Availability.** The simulation scripts used in this study are available from the corresponding author upon reasonable request.

**Competing Interests.** The authors declare no competing financial interest.

# Supporting Information

This Supporting Information provides additional details of the inverse-design methodology, electromagnetic simulations, fabrication-disorder modeling, quantum-protocol performance analysis, and statistical evaluation of the simulated yield. All device-performance results are obtained from three-dimensional finite-difference time-domain (FDTD) simulations. The investigated photonic components are optimized for the optical transitions of silicon T and A11 spin–photon centers; experimental device fabrication and quantum-protocol demonstrations are beyond the scope of this study.

**Contents**

## S1. Inverse-design framework and optimization convergence

Both devices were designed by gradient-based inverse design (the adjoint method)[1-3] on a silicon-on-insulator (SOI) platform with a 220 nm silicon device layer on a 2 µm silica buried-oxide layer ($n_{Si}$ = 3.46, $n_{SiO2}$ = 1.444), with air as the upper cladding and in the etched regions. The permittivity of a rectangular design region was represented by a level-set function and updated iteratively. Each iteration requires only two FDTD simulations, one forward and one adjoint, irrespective of the number of design degrees of freedom.

The objective functions encode the target function directly. For the Y-splitter, the objective maximizes the total power transmitted from the input waveguide into the fundamental ($TE_0$) mode of the two output waveguides while penalizing back-reflection into the input. For the directional coupler, it penalizes the deviations of the cross and bar powers from 0.5, the excess loss and the residual conversion into the first-order mode. The operating wavelength, 1326 nm for the T center and 1482.44 nm for the A11 center, enters as the wavelength at which the objective is evaluated. Explicitly, the Y-splitter, whose design region is mirror-symmetric by construction, minimizes $J_Y = 5|0.5 - P_{top}| + R + 0.01(f_{gap} + f_{curv})$, where $P_{top}$ is the $TE_0$ power in one output arm, $R$ the $TE_0$ back-reflection into the input, and $f_{gap}$ and $f_{curv}$ are normalized level-set gap and curvature penalties that target minimum feature sizes of 140 nm (Y-splitter) and 50 nm (directional coupler).[4] The directional coupler minimizes $J_{DC} = \Sigma_{IN1,IN2}[25(T_{cross} - 0.5)^2 + 25(T_{bar} - 0.5)^2 + 15(1 - \eta)^2 + 3T_{TE1}] + 0.05(f_{gap} + 1.5f_{curv})$, where $\eta = T_{cross} + T_{bar}$ and $T_{TE1}$ is the power converted into the first-order mode; by the mirror symmetry of the design, the IN2 terms equal the IN1 terms.

The access waveguides are 0.50 µm wide in the 0.22 µm silicon layer. The design regions are 1.7 × 1.7 µm for the Y-splitter and 13 × 2 µm for the directional coupler. The footprints quoted throughout are these optimization regions and exclude the input and output access waveguides. The design-region permittivity was defined on a 4 nm level-set grid for the Y-splitter and on a 6.7 nm (T center) or 7.5 nm (Al1 center) grid for the directional coupler. Figure S1 shows the convergence of the optimization objective −J, which the optimizer maximizes. The coupler objectives rise rapidly during the first 40 (Al1) to 60 (T) iterations and continue to improve slowly up to the 80-iteration limit, where the best designs are obtained. The Y-splitter objectives rise steeply over the first ~20 iterations and then continue to improve slowly up to the 150-iteration limit, with a residual gain of 0.10 (T center) and 0.05 (Al1 center) in objective units over the final 50 iterations. On the optimization grid, the final objective of the T-center Y-splitter ($J = 0.057$) implies $\eta \geq 1 - 2J/5 = 0.977$ at 1326 nm, whereas the validation grid gives 0.939 (Section S2); as for the coupler balance point (Section 2.2 of the main text), the T-center designs are the most sensitive to the grid refinement.

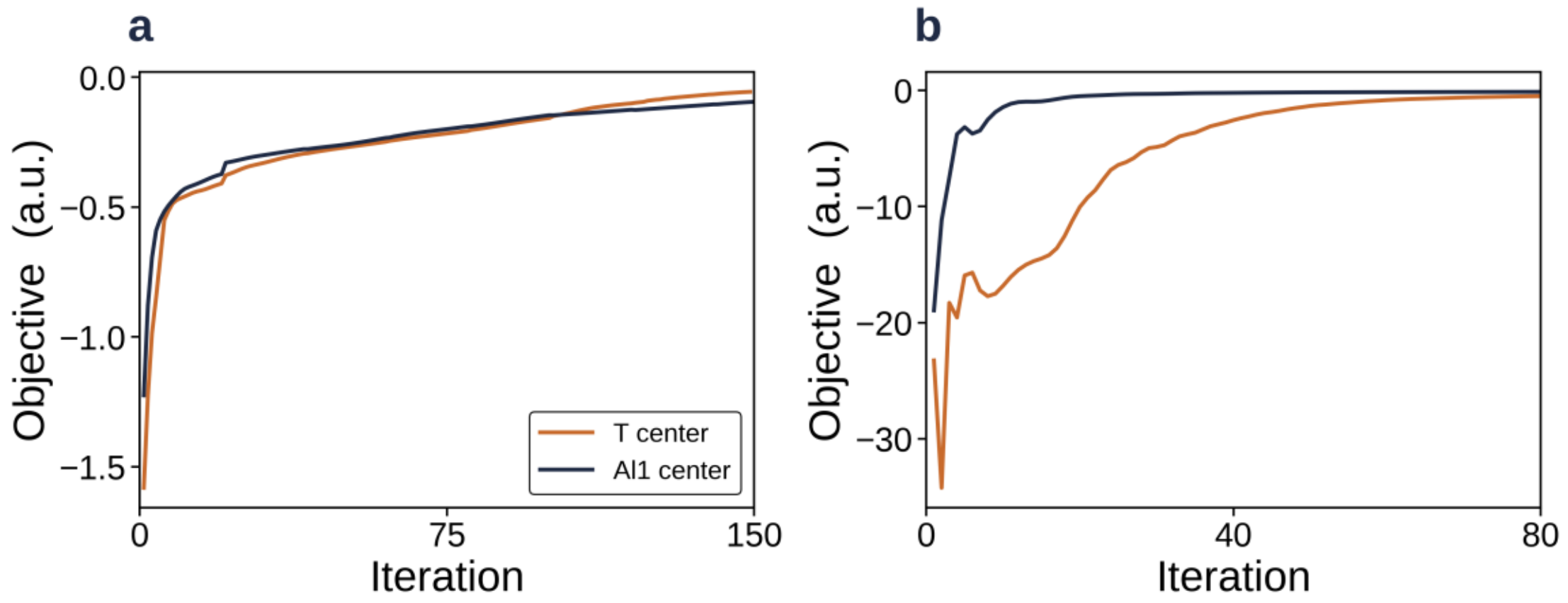

**Figure S1. Convergence of the adjoint inverse design.** Optimization objective −J (arbitrary units; higher is better) versus iteration for the (a) Y-splitter and (b) directional coupler optimized for the T center (orange) and the Al1 center (navy). Each iteration uses one forward and one adjoint FDTD simulation.

## S2. High-resolution electromagnetic validation and nominal device performance

Each optimized geometry was re-simulated on a Yee grid with a maximum cell size of $\lambda_c/(74\ n_{Si})$, i.e. 5.2 nm at 1326 nm and 5.8 nm at 1482.44 nm, with a 1.5 ps run time, for the Y-splitter, and with a maximum cell size of $\lambda_c/(57\ n_{Si})$ (6.7 and 7.5 nm) with a 5.0 ps run time for the directional coupler, with spectral sampling over ±100 nm (201 points) and ±50 nm (41 points), respectively, to obtain the nominal metrics reported in the main text at the spectral samples closest to the operating wavelengths (within 1 nm). The split ratio $r$ is the fraction of the transmitted power in one output arm (Y-splitter) or in the cross port (directional coupler). The total transmission $\eta$ is the sum of the $TE_0$ output-port powers (for the Y-splitter, whose arms are mirror images, twice the $TE_0$ power monitored in one arm), and the insertion loss IL = $-10 \log_{10} \eta$ is the excess loss beyond the ideal split. Table S1 lists these quantities, together with the back-reflection $R$ and, for the coupler, the residual leakage into the first higher-order guided mode (labelled $TE_1$), for both devices at both operating wavelengths.

| Device | λ (nm) | $r$ | $\eta$ | IL (dB) | $R$ | $TE_1$ leakage |
|---|---|---|---|---|---|---|
| Y-splitter, T | 1326 | 0.500 | 0.939 | 0.275 | $4.5\times10^{-4}$ | n/a |
| Y-splitter, Al1 | 1482.44 | 0.500 | 0.968 | 0.143 | $1.6\times10^{-4}$ | n/a |
| Dir. coupler, T | 1326 | 0.470 | 0.903 | 0.442 | $2.4\times10^{-5}$ | $5.0\times10^{-4}$ |

| Device | λ (nm) | *r* | *η* | IL (dB) | *R* | $TE_1$ leakage |
|---|---|---|---|---|---|---|
| Dir. coupler, Al1 | 1482.44 | 0.501 | 0.951 | 0.218 | $1.0\times10^{-5}$ | $2.7\times10^{-4}$ |

**Table S1. Nominal (high-resolution, simulated) performance of the two devices at the spectral samples closest to the T- and Al1-center operating wavelengths (Y-splitter: 1326.5 and 1482.7 nm; directional coupler: 1326.6 and 1483.3 nm).** Split ratio $r$, total transmission $\eta$, insertion (excess) loss IL, back-reflection $R$ and, for the directional coupler, residual $TE_1$ modal leakage. The cross and bar transmissions of the directional coupler are 0.425 and 0.479 (T center) and 0.476 and 0.475 (Al1 center). For the Y-splitter, $r = 0.500$ follows from the mirror symmetry of the design, and $\eta$ is twice the $TE_0$ power in one output arm (Section S3). Linear interpolation of the coupler spectra to the exact operating wavelengths gives r = 0.472 and 0.4985 and η = 0.904 and 0.952 for the T- and Al1-center couplers.

Figure S2 shows the same quantities across the full evaluation window. The transmission maxima lie close to, but not exactly at, the operating wavelengths. The T-center Y-splitter has a broad maximum near 1280 nm and retains $\eta = 0.939$ at 1326 nm. The Al1-center coupler peaks near 1455 nm. Because the objectives were evaluated at a single wavelength on the coarser optimization grid and also penalize small features (and, for the Y-splitter, back-reflection), the transmission maximum need not coincide with the operating wavelength on the validation grid. At the operating wavelengths the back-reflection is at most $4.5 \times 10^{-4}$ (T-center Y-splitter) and falls to the $10^{-5}$ level for the couplers. The coupler's $TE_1$ leakage remains at or below about $10^{-3}$ across the band. Both spurious channels are therefore negligible for cavity-fed operation.

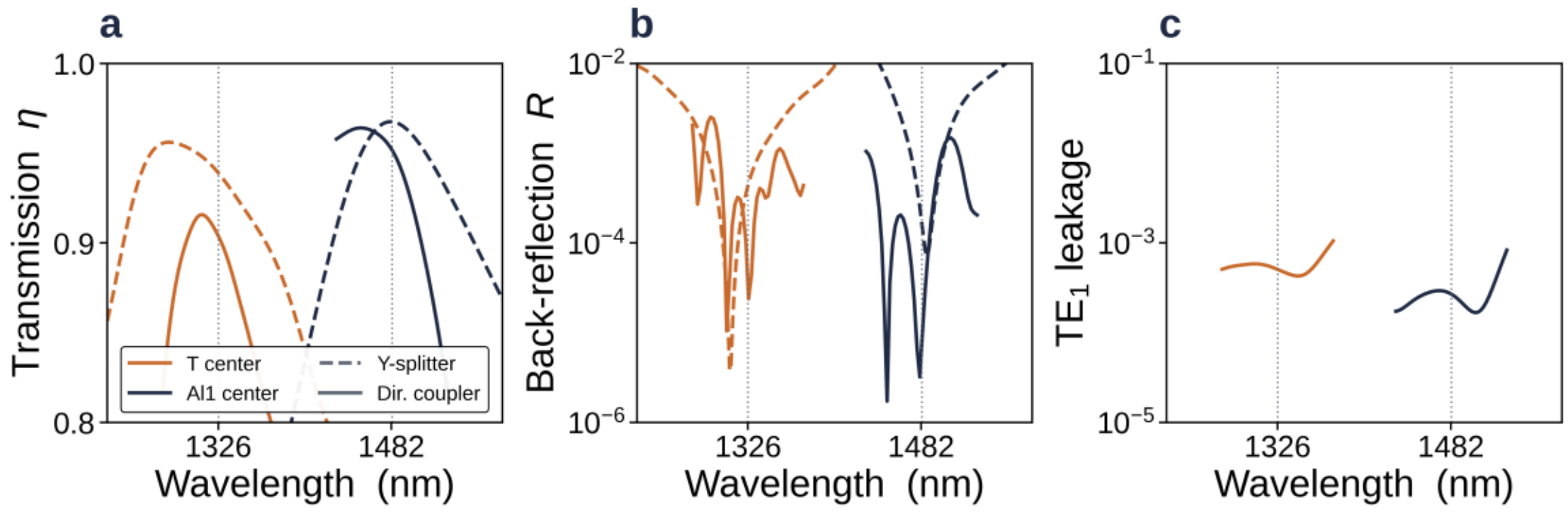

**Figure S2. Simulated S-parameter budget of the devices.** (a) Total transmission $\eta$, (b) back-reflection $R$ and (c) residual $TE_1$ leakage of the directional coupler, versus wavelength, for the Y-splitter (dashed) and directional coupler (solid) optimized for the T center (orange) and Al1 center (navy). Dotted vertical lines mark the operating wavelengths. The Y-splitter is shown over a ±100 nm window and the coupler over a ±50 nm window.

## S3. Fabrication-tolerance analysis and variance budget

**Disorder model.** Robustness to fabrication imperfection was assessed by a Monte Carlo study of 100 independent disorder realizations for each device and each operating wavelength. Every realization applied four independent perturbations, with magnitudes chosen to represent local, within-die variations (the LER follows reported silicon-on-insulator statistics,[5] and the other terms are representative values for silicon photonic processes[6, 7]):

1. a global etch bias, i.e. a uniform dilation or erosion of all feature boundaries (1σ = 2 nm);
2. line-edge roughness (LER), a spatially correlated perturbation of the boundaries (3 nm rms, 50 nm correlation length);
3. a variation of the silicon-film thickness (1σ = 3 nm);
4. a variation of the silicon refractive index (1σ = 0.005).

The etch bias and LER are implemented as a uniform shift and a correlated random field, respectively, of the level-set function. Uniform boundary shifts of this kind (erosion and dilation) are commonly used to design for, and to assess, the fabrication tolerance of inverse-designed devices.[2, 8, 9] The four ensembles use the same fixed random seed for reproducibility, and each was re-simulated on the high-resolution grid to obtain $r$, $\eta$ and IL. This is a statistical ensemble, not a worst-case corner analysis. The ensemble simulations use the same spatial grid as the Table S1 validation and a run time of 1.0 ps, and each ensemble is referenced to a nominal simulation computed with identical settings, which gives $r = 0.4952$ and $\eta = 0.9496$ (T-center Y-splitter), 0.5003 and 0.9665 (Al1-center Y-splitter), 0.4740 and 0.9043 (T-center coupler), and 0.4988 and 0.9518 (Al1-center coupler). For the Y-splitter, the ensemble metrics are evaluated at the closest sampled wavelengths, 1324.1 and 1480.8 nm. The T-center Y-splitter ensemble is centered at $r = 0.495$, in agreement with its nominal at the same settings: the ensemble mean differs from it by 0.1 standard error, but from 0.5 by about 21 standard errors. Because the geometry is mirror-symmetric to machine precision, etch bias, thickness and index preserve this symmetry, and the LER has zero mean, the offset is numerical. It originates in the interpolation of the design-region permittivity onto the FDTD grid. The level-set permittivity is sampled every 4 nm, at odd multiples of 2 nm from the device axis, and each grid node takes the value of the nearest sample. On the T-center grid, the 100 nm intervals between the waveguide edges at $|y| = 0.15$ and 0.25 μm are divided into cells of exactly 5.0 nm, which places grid nodes at multiples of 4 nm, exactly midway between two samples. In three of these node rows ($|y| = 0.16$, 0.22 and 0.24 μm), the nearest sample is taken on the same ($+y$) side in both arms, so that 90 mirror pairs of cells in the junction (0.17% of the design-region cells) differ in permittivity. The grid, the static structures, the source and the monitors are otherwise mirror-symmetric, so this sampling asymmetry is the only asymmetry of the simulated structure and the origin of the offset. The in-plane field of the validation simulation reproduces it ($r = 0.4953$ from the field intensity in the two output arms). On the Al1-center grid, the same intervals contain 5.56 nm cells, fewer nodes fall midway between samples, and the offset is about 15 times smaller ($r = 0.5003$). The offset lowers the min-entropy yield of the raw T-center ensemble at the 0.99 requirement to 26%. Because the split-ratio spread is set by the LER and is unaffected by this offset (the ensemble mean differs from the nominal by 0.1 standard error), the yield referenced to the balanced nominal, 90% (95% confidence interval 82.4–95.1%), is the physically relevant value. The same offset explains most of the difference between the transmission of the ensemble nominal ($\eta = 0.9496$) and that of Table S1 ($\eta = 0.939$): Table S1 doubles the $TE_0$ power of the top arm, which the offset makes the weaker one. The in-plane field of the validation simulation gives a bottom-to-top power ratio of 1.019, so that the two-arm sum on the validation grid is 0.948; the small remainder comes from the ensemble settings (spectral sample at 1324.1 nm instead of 1326.5 nm and 1.0 ps run time).

**Statistics.** Table S2 summarizes the distributions. The Y-splitter split ratio remains extremely tight ($\sigma_r$ = 0.0023 to 0.0034). The directional coupler shows a wider distribution, centered on its own nominal split ratio (means 0.475 and 0.498 for the T- and Al1-center couplers, $\sigma_r$ = 0.020 to 0.024), and every realization stays within about 0.07 of the ideal $r = 0.5$.

| Device | $r$ (mean ± sd) | $\eta$ (mean ± sd) | IL (dB) | $F_{min}$ | $V_{min}$ | $H_{min}$ (worst, bit) |
|---|---|---|---|---|---|---|
| Y-splitter, T | 0.495 ± 0.002 | 0.948 ± 0.004 | 0.233 ± 0.017 | >0.9998 | n/a[a] | 0.967 |
| Y-splitter, Al1 | 0.500 ± 0.003 | 0.965 ± 0.002 | 0.157 ± 0.009 | >0.9998 | n/a[a] | 0.974 |
| Dir. coupler, T | 0.475 ± 0.020 | 0.901 ± 0.008 | 0.453 ± 0.039 | >0.994 | 0.957 | 0.801 |
| Dir. coupler, Al1 | 0.498 ± 0.024 | 0.948 ± 0.010 | 0.233 ± 0.044 | >0.994 | 0.958 | 0.803 |

**Table S2. Fabrication-tolerance statistics from the 100-realization Monte Carlo ensemble.** Split ratio $r$, total transmission $\eta$ and insertion loss IL are given as mean ± standard deviation. $F_{min}$, $V_{min}$ and $H_{min}$ (worst) are the minima over the 100 realizations of the dual-rail fidelity, the HOM visibility and the random-number min-entropy. Including the nominal coupler phase (Section S4), the minimum visibilities of the couplers become 0.956 (T) and 0.957 (Al1). The ensemble statistics are computed with the ensemble settings of Section S3, whose nominal values differ slightly from those of Table S1; for the T-center Y-splitter ($\eta$ = 0.9496 instead of 0.939), most of the difference arises because Table S1 doubles the power of one arm (Section S3). The T-center Y-splitter statistics include the numerical offset of Section S3; referenced to the balanced nominal, its mean split ratio is 0.500 and its worst-case min-entropy is 0.981 bit. [a]A 1×2 splitter has a single input and does not support two-photon interference. Its formal value $V(r) > 0.998$ is plotted in Figure S4b only as a measure of balance.

**Per-source sensitivity and variance budget.** A standardized multiple-linear regression of $r$ on the four disorder sources (Table S3; see also Figure 3d of the main text) identifies the dominant contributions. Each entry is $|\beta_i| \times \mathrm{sd}(r)$, the split-ratio shift produced by a one-standard-deviation change of source $i$. Because the sources are independent, the quadrature sum of the entries approximately equals $\sqrt{(R^2)} \times \mathrm{sd}(r)$, where $R^2$ is the coefficient of determination of the linear model.

| Device | Etch bias | LER | Si thickness | Refr. index | Quadrature sum | $\sigma_r$ | $R^2$ |
|---|---|---|---|---|---|---|---|
| Y-splitter, T | 0.0001 | 0.0001 | 0.0003 | 0.0004 | 0.0005 | 0.002 | 0.055 |
| Y-splitter, Al1 | 0.0000 | 0.0005 | 0.0002 | 0.0006 | 0.0008 | 0.003 | 0.041 |
| Dir. coupler, T | 0.0105 | 0.0005 | 0.0085 | 0.0145 | 0.0198 | 0.020 | 0.943 |
| Dir. coupler, Al1 | 0.0036 | 0.0003 | 0.0164 | 0.0171 | 0.0240 | 0.024 | 0.982 |

**Table S3. Per-source split-ratio sensitivity and variance budget from the Monte Carlo ensemble.** Entries are the split-ratio shift induced by a one-standard-deviation change of each disorder source. The quadrature sum is computed from the tabulated values, and $R^2$ is the coefficient of determination of the unrounded regression. The Y-splitter is insensitive to all four scalar sources. The directional coupler is dominated by refractive-index and etch-bias variation at 1326 nm, and by silicon-thickness and refractive-index variation at 1482.44 nm.

For the directional coupler the four sources account for 94% (T) and 98% (Al1) of the split-ratio variance, so a linear, effective-index-driven model captures nearly all of it. At the T-center wavelength the coupler responds most strongly to refractive-index and etch-bias variation; at the Al1-center wavelength it responds to thickness and refractive-index variation. For the Y-splitter, etch bias, film thickness and refractive index preserve the mirror symmetry and therefore cannot change $r$, consistent with their near-zero coefficients. The residual $\sigma_r \approx 0.002$ to 0.003 arises from the symmetry-breaking spatial pattern of the LER, which a scalar roughness amplitude does not capture, hence the low $R^2$. Figure S3 shows the Monte Carlo split-ratio spectral envelopes and the insertion-loss distributions.

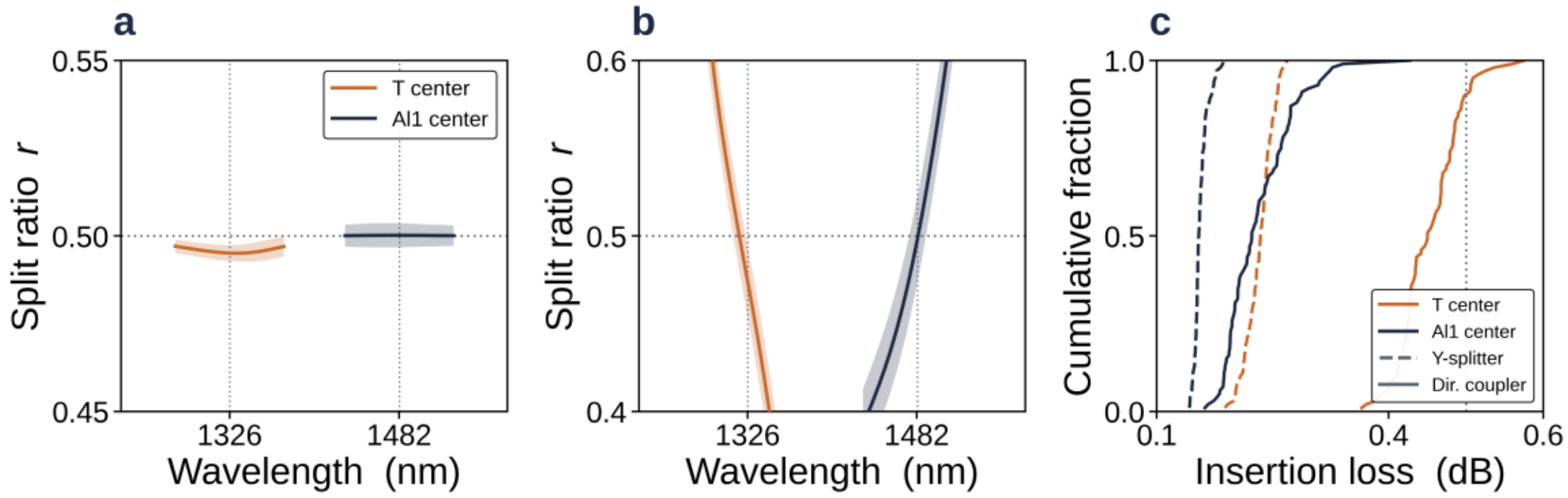

**Figure S3. Monte Carlo fabrication-tolerance response.** Split-ratio spectral envelope (mean, solid line; ±1 standard deviation, shaded band) over the 100 realizations for the (a) Y-splitter and (b) directional coupler at the T center (orange) and Al1 center (navy); the horizontal line marks $r = 0.5$. (c) Cumulative distribution of the insertion loss over the 100 realizations for the Y-splitter (dashed) and directional coupler (solid); the dotted line marks a 0.5 dB reference.

## S4. Quantum-protocol figures of merit, analytical extensions, and simulated yield

**Assumptions.** At its operating wavelength, each device is modeled as a beam-splitting element of power split ratio $r$. Its insertion loss is factored out: in heralded and post-selected protocols, balanced loss reduces the success rate rather than the conditional fidelity (Section S5). Unless stated otherwise, the interfering photons are indistinguishable and the detectors ideal and identical.

**Dual-rail state fidelity.** A single photon incident on the splitter is transformed into $|\psi\rangle = \sqrt{r}\,|1,0\rangle + e^{i\theta}\sqrt{(1-r)}\,|0,1\rangle$. The phase $\theta$ is a static, calibratable property of the device and is compensated by a phase shifter. The fidelity to the target state $|+\rangle = (|1,0\rangle + |0,1\rangle)/\sqrt{2}$, the dual-rail $|+\rangle$ qubit and the maximally entangled single-photon (mode-entangled) state,[10, 11] is then

$$F(r) = \frac{1}{2}\left(\sqrt{r} + \sqrt{1-r}\right)^2 = \frac{1}{2} + \sqrt{r(1-r)}. \qquad \text{(S1)}$$

**HOM visibility of a lossless coupler.** For two photons entering opposite ports of a lossless 2×2 coupler with power reflectivity $1 - r$ and transmissivity $r$, the coincidence probability is $(1 - 2r)^2$ for indistinguishable photons and $r^2 + (1 - r)^2$ for distinguishable ones.[12, 13] The visibility, $V = 1 - P_{c,\text{indist}}/P_{c,\text{dist}}$, is

$$V(r) = \frac{2r(1-r)}{r^2 + (1-r)^2}. \qquad \text{(S2)}$$

**Extension 1: lossy coupler and cross–bar phase.** For a lossy coupler, unitarity no longer fixes the relative phase of the cross and bar amplitudes.[14, 15] Consider a coupler that is mirror-symmetric about its axis, with complex bar amplitude $b$ and cross amplitude $c$ for either input. The two-photon coincidence amplitude is $b^2 + c^2$, so

$$V = -\frac{2|b|^2|c|^2\cos[2(\phi_b - \phi_c)]}{|b|^4 + |c|^4} = V(r)\cos(2\,\Delta\phi), \qquad \text{(S3)}$$

where $r = |c|^2/(|b|^2 + |c|^2)$ and $\Delta\phi = \phi_b - \phi_c - \pi/2$ is the departure from phase quadrature.[15] A lossless coupler has $\Delta\phi = 0$ and recovers eq S2. Equation S2 therefore gives the visibility for ideal phase, and eq S3 the general result. Both $\phi_b$ and $\phi_c$ follow from the complex mode amplitudes of the single-input simulations reported here. For the nominal couplers, the complex $TE_0$ amplitudes of the two single-input validation simulations give $\Delta\phi = -1.5°$ (T center) and $-1.0°$ (Al1 center), well within the passivity bound $|\sin \Delta\phi| \leq (1 - \eta)/(2|b||c|)$, i.e. 6.2° and 3.0°.[15] The factor $\cos 2\Delta\phi = 0.9987$ and 0.9993 lowers the device-limited visibility from 0.9929 to 0.9916 (T) and from 0.99999 to 0.9993 (Al1). The full 2×2 scattering matrix reproduces these values to within $10^{-5}$, the nominal devices being mirror-symmetric to $10^{-6}$. Applying the nominal phase to the Monte Carlo ensembles leaves the 0.95 visibility yields at 100% and changes the 0.99 yields from 68% to 64% (T) and from 91% to 89% (Al1), within the binomial confidence intervals of Table S5. LER breaks the mirror symmetry, so a rigorous per-realization evaluation would require the full 2×2 scattering matrix, with the second input excited in one additional simulation per realization; the general coincidence amplitude is then $S_{31}S_{42} + S_{32}S_{41}$. Per-realization phases were not computed; the nominal phases lie well within the passivity bound, and the heralded Barrett–Kok fidelity does not depend on this phase (Extension 3).

**Extension 2: partial distinguishability.** If the two photons have mode overlap $M$ ($0 \leq M \leq 1$), the indistinguishable-photon interference term is weighted by $M$, giving

$$V = M\,V(r)\cos(2\,\Delta\phi). \qquad \text{(S4)}$$

The device-limited visibility reported in the main text is therefore an upper bound that multiplies the emitter indistinguishability.

**Extension 3: Barrett–Kok heralding with an unbalanced splitter.** Consider two ideal emitters whose spin–photon states, $(|\uparrow\rangle|1\rangle + |\downarrow\rangle|0\rangle)/\sqrt{2}$, are interfered on a lossless splitter of split ratio $r$, followed by a spin flip and a second round, as in the Barrett–Kok protocol.[16] When both heralding clicks occur in the same detector, the two spin configurations acquire the same amplitude product over the two rounds, $\sqrt{[r(1-r)]}$, and the heralded spin state is maximally entangled for any $r$. When they occur in opposite detectors, the heralded state is proportional to $r|\uparrow\downarrow\rangle + e^{i\chi}(1-r)|\downarrow\uparrow\rangle$, where the phase $\chi$ is fixed by the device and can be corrected, and its fidelity to the target Bell state is $F_{opp} = 1/(2[r^2 + (1-r)^2]) = [1 + V(r)]/2$. The two cases occur with probabilities $2r(1-r)$ and $r^2 + (1-r)^2$, so the average heralded fidelity is $1 - 2\delta^2$, with $\delta = r - 0.5$. A device-limited visibility $V(r) = 0.99$ therefore guarantees $F \geq 0.995$, with an average of 0.9975. When $V$ is instead limited by photon distinguishability, the same expression, $(1 + V)/2$, bounds the heralded fidelity from above for a balanced splitter.[17, 18]

**Random-number min-entropy.** In a branching-path random-number generator, each detected photon yields the bit 0 with probability $r$ and the bit 1 with probability $1 - r$. The min-entropy per bit is[19]

$$H_{\min}(r) = -\log_2[\max(r,\, 1-r)]. \qquad \text{(S5)}$$

Unequal detector efficiencies $d_1$ and $d_2$ replace $r$ by $r' = rd_1/[rd_1 + (1-r)d_2]$, so detector matching matters as much as splitter balance. A residual bias is removed by randomness extraction, for example by Toeplitz hashing, which compresses $n$ raw bits to about $n\,H_{\min}$ nearly uniform bits.[20] $H_{\min}$ therefore sets the extraction ratio rather than a hard pass/fail threshold.

**Sensitivity to imbalance.** Each figure of merit reaches its ideal value ($F = V = 1$, $H_{\min} = 1$ bit) at $r = 0.5$. To leading order in $\delta = r - 0.5$, $F \approx 1 - \delta^2$ and $V \approx 1 - 8\delta^2$ are quadratic, whereas $H_{\min} \approx 1 - 2|\delta|/\ln 2$ has a linear cusp. Consequently, a requirement of 0.95 tolerates $|\delta| \leq 0.218$ for $F$, $|\delta| \leq 0.080$ for $V$ and only $|\delta| \leq 0.0176$ for $H_{\min}$. At 0.99 the tolerances are 0.0995, 0.0354 and 0.0035, respectively.

**Complete yield curves.** Figure S4 gives the simulated yield, the fraction of the 100 realizations meeting a given requirement, for all three figures of merit and all four devices. The dual-rail fidelity yield is 100% at a 0.99 requirement for both devices. The coupler's HOM visibility yield is 100% at 0.95. The min-entropy, with its cusp, is the most demanding. The Y-splitter retains 100% yield at 0.95 bit, whereas the coupler meets that requirement in 36% (T) and 53% (All) of realizations. The coupler is therefore best suited to the two-photon role for which it is built, and the symmetry-protected Y-splitter to the entropy-limited role.

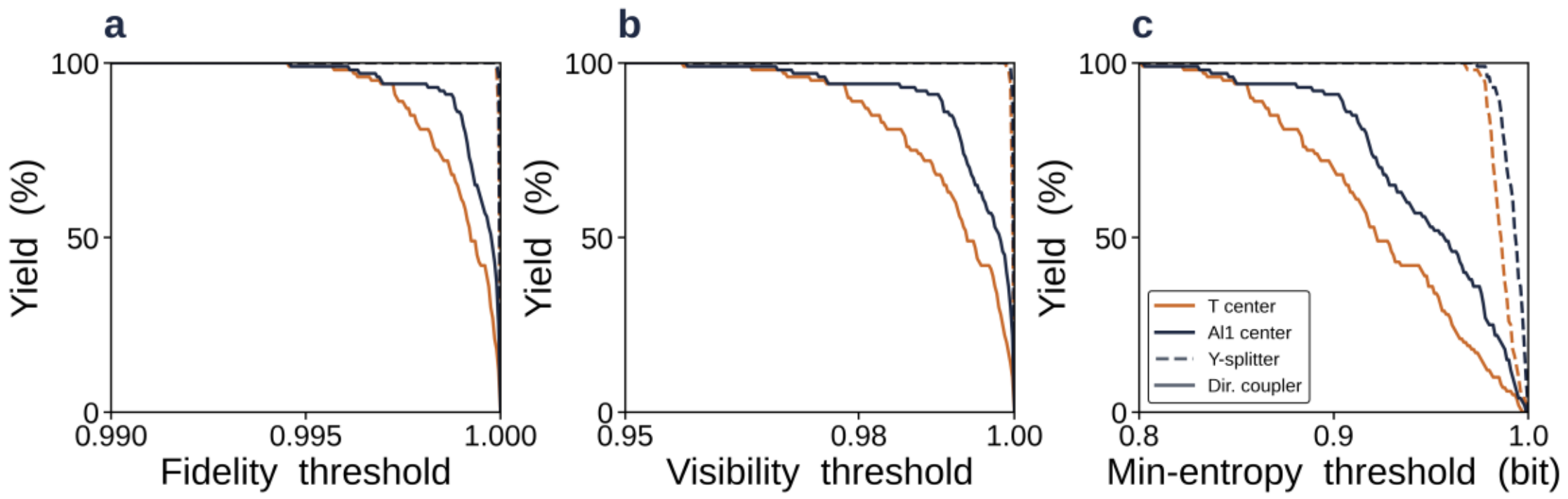

**Figure S4. Complete protocol-level simulated yield.** Yield, the fraction of the 100 Monte Carlo realizations meeting a given protocol requirement, versus that requirement for (a) dual-rail fidelity, (b) HOM visibility and (c) random-number min-entropy. Curves are shown for the Y-splitter (dashed) and directional coupler (solid) at the T center (orange) and Al1 center (navy). The Y-splitter curves in (b) are the formal $V(r)$ and are shown only as a measure of balance (see Table S2, note a). The T-center Y-splitter curve in (c) is for the raw ensemble, which carries the numerical offset of Section S3; referenced to the balanced nominal, its yield at 0.99 bit is 90%.

## S5. Back-reflection, optical loss, and integration considerations

Each device is characterized in isolation with a waveguide-mode source, the standard approach for inverse-designed passive components. Three device-level quantities determine how this characterization carries over to an integrated link.

**Transmission.** A well-designed cavity-coupled emitter feeds the fundamental guided mode of the access waveguide, which is the mode for which the device is optimized. The transmission $\eta$ reported here is therefore expected to be the relevant integrated quantity at leading order. In single-photon protocols $\eta$ scales the count rate. In two-photon coincidence protocols, such as HOM interference and Barrett–Kok heralding, the success probability scales as $\eta^2$, which is 0.82 and 0.90 for the T- and Al1-center couplers; for the Y-splitters, $\eta^2 = 0.88$ and 0.94 applies when two photons must each traverse a splitter. Balanced loss reduces the rate but not the heralded fidelity.

**Back-reflection.** The back-reflection $R$ of each device (Table S1) is at most $4.5 \times 10^{-4}$, so reflection back into the emitter cavity or any upstream element is negligible. Reflection would matter only for cavities of very high finesse, where light returning into the cavity could perturb the emission; such cavities are not considered here.

**Integration.** A full cavity-to-device coupled simulation, including the cavity–waveguide transition, is left to future work, together with fabrication and measurement.

## S6. Extended computational methods and simulation parameters

All electromagnetic simulations used the Tidy3D FDTD solver (Flexcompute Inc.) with adjoint (automatic-differentiation) gradients. Table S4 collects the numerical parameters required to reproduce the results.

| Parameter | Value |
|---|---|
| Device layer / substrate / cladding | 220 nm Si ($n$ = 3.46) / $SiO_2$ ($n$ = 1.444, 2 µm) / air |
| Access waveguides | 0.50 µm wide, $TE_0$ excitation |

| Parameter | Value |
| --- | --- |
| Design region | 1.7 × 1.7 µm (Y-splitter); 13 × 2 µm (directional coupler) |
| Parametrization | Level set; 4 nm grid (Y-splitter), 6.7/7.5 nm grid (coupler) |
| Optimizer, step size, iterations | Adam with adaptive step size, 0.3 → 0.01 (Y-splitter) and 0.15 → 0.05 (coupler); up to 150 (Y-splitter), 80 (coupler) iterations |
| Minimum feature size / radius of curvature | Targets of 140 nm (Y-splitter) and 50 nm (coupler), soft level-set gap and curvature penalties; final Y-splitter air features down to about 60 nm |
| Optimization grid | 37 (Y-splitter) and 45 (coupler) points per wavelength in Si, i.e. ≈10.4/11.6 nm and ≈8.5/9.5 nm |
| Validation grid | Maximum cell size $\lambda_c/(74\ n_{Si})$: 5.2 nm (1326 nm), 5.8 nm (1482.44 nm), Y-splitter; $\lambda_c/(57\ n_{Si})$: 6.7 nm (1326 nm), 7.5 nm (1482.44 nm), coupler |
| Run time (validation / ensemble) | 1.5 ps (Y-splitter), 5.0 ps (coupler) / 1.0 ps |
| Boundary conditions | PML (12 layers) on all sides, 0.6 $\lambda_c$ from the structures |
| Sources and monitors | $TE_0$ mode source; mode monitors on all ports; coupler validation with each input excited separately |
| Monte Carlo ensemble | 100 realizations per device and wavelength; same fixed seed for all ensembles |

**Table S4. Numerical parameters.** Parameters of the optimization, high-resolution validation and Monte Carlo simulations.

## S7. Statistical uncertainty of Monte Carlo yield estimates

With $N = 100$ realizations, the standard error of a mean split ratio is $\sigma_r/\sqrt{N}$, i.e. 0.0002 to 0.0003 for the Y-splitter and 0.002 for the coupler. The relative standard error of a standard deviation is approximately $1/\sqrt{[2(N-1)]} \approx 7\%$. Yields are binomial proportions. Table S5 gives exact (Clopper–Pearson) 95% confidence intervals[21] for the principal yields quoted in the main text. A yield of 100/100 has a lower 95% confidence bound of 96.4%. Narrowing the yield intervals at the 0.99 requirement to about ±3 percentage points would require roughly 1,000 realizations per device.

| Device and figure of merit | Requirement | Yield (k/100) | 95% CI (%) |
| --- | --- | --- | --- |
| Y-splitter, $H_{min}$ (T, Al1) | 0.95 bit | 100, 100 | 96.4–100 |
| Dir. coupler, $V$ (T, Al1) | 0.95 | 100, 100 | 96.4–100 |
| Dir. coupler, $V$ (T) | 0.99 | 68 | 57.9–77.0 |
| Dir. coupler, $V$ (Al1) | 0.99 | 91 | 83.6–95.8 |
| Y-splitter, $H_{min}$ (T) | 0.99 bit | 26 (raw), 90 (balanced) | 17.7–35.7, 82.4–95.1 |
| Y-splitter, $H_{min}$ (Al1) | 0.99 bit | 66 | 55.8–75.2 |
| Dir. coupler, $H_{min}$ (T) | 0.95 bit | 36 | 26.6–46.2 |
| Dir. coupler, $H_{min}$ (Al1) | 0.95 bit | 53 | 42.8–63.1 |

**Table S5. Simulated yields with exact binomial 95% confidence intervals.** Including the nominal coupler phase (Section S4), the 0.99 visibility yields become 64/100 (95% CI 53.8–73.4%) for the T center and 89/100 (81.2–94.4%) for the Al1 center. For the T-center Y-splitter at 0.99 bit, the first value is for the raw ensemble, which carries the numerical offset of Section S3, and the second is referenced to the balanced nominal.

## Supporting References